\documentclass[%
 reprint,
 amsmath,amssymb,
 aps,
]{revtex4-2}

\usepackage{graphicx}
\usepackage{dcolumn}
\usepackage{bm}
\usepackage{orcidlink}

\usepackage{xcolor}

\begin{document}

\preprint{APS/123-QED}

\title{Third harmonic generation in the transition metal trichalcogenides}

\author{Timothy Y. Chung}
\author{Vergil M. Schreiber}
\author{Randy M. Sterbentz}
\author{David P. Shelton}
\author{Joshua O. Island}%
\email{joshua.island@unlv.edu}
\affiliation{Department of Physics and Astronomy, University of Nevada Las Vegas, Las Vegas, NV 89154, USA
}%

\author{Eduardo Flores}
\affiliation{Departamento de Física Aplicada, Centro de Investigación y Estudios Avanzados, 97310 Mérida, México}%

\author{Jose Ramon Ares}
\author{Isabel J. Ferrer}
\altaffiliation[Also at ]{Instituto Nicolás Cabrera (INC), Universidad Autónoma de Madrid, UAM, 28049 Madrid, Spain}
\affiliation{Departamento de Física de
Materiales, Universidad Autónoma de Madrid, UAM, 28049 Madrid, Spain}%

\date{\today}

\begin{abstract}
The transition metal trichalcogenides (TMTs) are an interesting class of layered 2D materials that have strong anisotropic electrical and optical properties. We report on the polarization-dependent nonlinear optical response of exfoliated crystals at the third harmonic using non-resonant excitation with a 1560~nm femtosecond laser. All members of the TMT crystal family explored (ZrS$_3$, HfS$_3$, TiS$_3$, and NbS$_3$) display large anisotropy in their polarized third harmonic generation signal, with anisotropy ratios as large as $\chi_{11}/\chi_{22} \approx 6.5$. From the expected nonlinear dependence of these monoclinic and triclinic crystals we extract the relative nonlinear susceptibility coefficients. We find that the anisotropy ratio is largest for ZrS$_3$ and decreases across the family, a trend we associate with the proximity of the third harmonic energy to the strongly $b$-polarized absorption edge of each material. NbS$_3$, the only triclinic crystal, displays a pronounced anisotropy in thick flakes that is strongly reduced in the
thinnest flakes, which we attribute to a partial relaxation of the Peierls-type Nb--Nb dimerization that underlies the triclinic distortion. The large anisotropic nonlinear response in this crystal family sets it apart from other 2D materials and makes it particularly attractive for further investigation.
\end{abstract}

\maketitle


\section{\label{sec:level1}Introduction}
Reduced crystal symmetries often lead to novel phenomena, functionality, and applications. This is exemplified in the layered, van der Waals materials like graphite and the transition metal dichalcogenides (TMDs). These materials are composed of covalently bonded sheets that are stacked through van der Waals forces to create 3D crystals. While the sheets are structurally symmetric in-plane, the reduced crystal symmetry in the out-of-plane direction is precisely why these materials can be exfoliated and isolated down to single monolayers. Mono- and few-layer flakes of these materials, and similar layered crystals, have generated a tremendous amount of interest and spawned an entirely new field of research in materials science focused solely on 2D materials. 

While most 2D materials to date possess largely isotropic in-plane electrical and optical properties, a few exceptions have emerged. Black phosphorus is perhaps the most well-known of the anisotropic 2D materials\cite{xia2014rediscovering, castellanos2014isolation}. Its in-plane puckered crystal structure results in linear, Dirac-like band structure along one direction and quadratic, Schr\"{o}dinger-like band structure along the other\cite{margot2023electronic}. This leads to anisotropic conductance\cite{xia2014rediscovering, liu2014phosphorene}, as well as linear dichroism\cite{yuan2015polarization}, and direction-dependent thermal\cite{lee2015anisotropic} and mechanical\cite{chen2016anisotropic, jiang2014mechanical, galluzzi2020mechanical} properties. This broken in-plane rotational symmetry also has significance in fundamental studies of the fractional quantum Hall effect where anisotropic bubble and stripe phases exist\cite{yang2018integer, fogler2002stripe}. A few other notable anisotropic 2D materials have been (re)discovered including the group IV monochalcogenides (MX, M: Ge, Sn, X: S, Se)\cite{chen2022plane, zhou2018highly}, the 1T$'$ or Td TMD semimetals\cite{kang2019nonlinear}, and the rhenium dichalcogenides.\cite{lin2015single} A wandering principal axis was recently discovered in the rhenium dichalcogenides that leads to wavelength-switchable propagation directions of its waveguide modes\cite{ermolaev2024wandering}.

\begin{figure}
\includegraphics[width=0.5\textwidth]{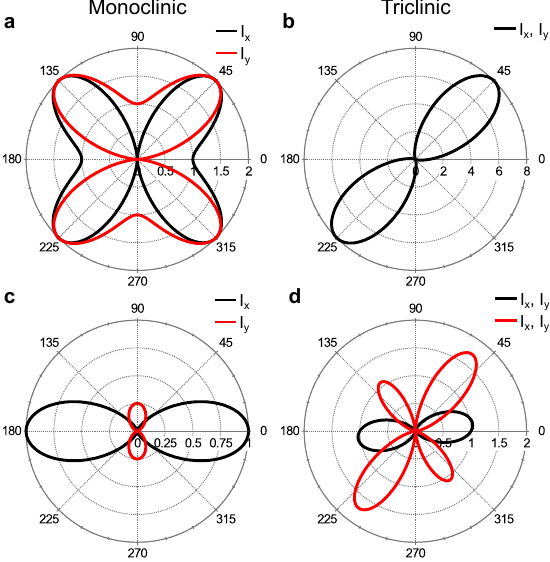}
\caption{\label{fig1} Expected polarization dependence for monoclinic and triclinic crystals. (a) Third harmonic intensity as a function of polarization for a monoclinic crystal with all susceptibility coefficients equal to 1. (b) Third harmonic intensity as a function of polarization for a triclinic crystal with all susceptibility coefficients equal to 1. For this case, $I_x$ and $I_y$ are equal for triclinic crystals with normal laser incidence. (c) Third harmonic intensity as a function of polarization for a monoclinic crystal with susceptibility coefficients $\chi_{11}=1, \chi_{18}=0.1$ (black horizontal two-lobe pattern) and $\chi_{22}=0.5, \chi_{29}=0.1$ (red vertical two-lobe pattern). (d) Third harmonic intensity as a function of polarization for a triclinic crystal with all susceptibility coefficients suppressed (0.1) except one (black two-lobe pattern with $\chi_{11}=\chi_{22}=1$, red four-lobe pattern with $\chi_{18}=\chi_{28}=1$). }
\end{figure}

The transition metal trichalcogenides (TMTs) are an interesting class of layered materials when compared to the aforementioned 2D materials because the entire family possesses reduced in-plane structural symmetry and the members span a spectrum of electronic phases (metals, superconductors, and semiconductors)\cite{island2017electronics, patra2020anisotropic, chen2023quasi}. Sometimes referred to as ``quasi-1D 2D materials'', the TMTs are composed of 1D prismatic chains of transition metal atoms surrounded by chalcogen atoms that are covalently bonded in-plane. The transition metal atom distances are slightly smaller along the chains than across leading to anisotropies in electrical and optical properties. Island \textit{et al.} have shown that exfoliated TiS$_3$ displays in-plane mobility ratios as high as 7.6 when comparing electrical transport along the a- and b-axes of the crystal and optical transmission ratios up to 30 in the visible wavelengths\cite{island2016titanium}. This conductance anisotropy in TiS$_3$ is larger than that found for black phosphorus flakes\cite{island2016titanium}. Optical measurements of the complex index of refraction of TiS$_3$ have also shown that the exfoliated crystals are strongly birefringent with magnitudes ($\delta n = +0.30 \pm 0.04$) larger than the well-known birefringent materials TiO$_2$ and calcite\cite{papadopoulos2018large}.

Polarization-dependent harmonic generation is a powerful tool in the study of 2D materials\cite{khan2022optical}. It has been used to investigate crystal symmetries, layer number, strain, defects, and stacking orientations in van der Waals heterostructures. Due to its sensitivity to all these aspects of 2D crystals, it also provides an important tool in the careful study of phase transitions. Anisotropic third harmonic generation has been reported in other low-symmetry 2D materials, including GeSe\cite{dasgupta} and SnS\cite{maragkakis}, and strain-tunable THG anisotropy has been demonstrated in Nb$_2$SiTe$_4$~\cite{ouyang}, placing the trichalcogenides within a broader family of materials with strong nonlinear optical anisotropy. In the TMT family, third harmonic generation has recently been reported in ZrS$_3$ and TiS$_3$~\cite{zhu2025orientation,zhang2026efficient}. Zhu \textit{et al.} found a maximum third-order susceptibility $\chi^{(3)} = 67.9\times10^{-18}$~m$^2$/V$^2$ for a 24.3~nm ZrS$_3$ flake and an intensity anisotropy ratio as high as 61.5 for a 17.7~nm flake\cite{zhu2025orientation}. Zhang \textit{et al.} reported efficient THG in TiS$_3$, characterizing its intrinsic polarization anisotropy, wavelength dependence, and the strain tuning of that anisotropy\cite{zhang2026efficient}. These reports motivate a systematic comparison of the nonlinear response across the crystal family under identical measurement conditions.

Here we investigate the third-order nonlinear optical properties of four members of the TMTs, ZrS$_3$, HfS$_3$, TiS$_3$, and NbS$_3$, measured on a single setup. To our knowledge, THG has not previously been reported in HfS$_3$ or NbS$_3$. Measuring all four crystals under identical conditions, we identify a systematic ordering of the anisotropy across the family and quantify the overall ratio. In NbS$_3$ we find a transition from strong anisotropy in thick flakes toward equal-amplitude lobes in the thinnest flakes, in contrast to the other three members, whose extracted susceptibility coefficients are largely thickness-independent. We hypothesize that this thickness dependence arises from a partial relaxation of the Peierls-type Nb--Nb dimerization that underlies the triclinic distortion of NbS$_3$, driving the thinnest flakes toward the undimerized, ZrS$_3$-like limit. 

\section{\label{sec:level1}Third harmonic generation in monoclinic and triclinic crystals}

When subjected to intense optical fields, certain materials will present a nonlinear response in their polarization. The third-order nonlinear polarization is written as:
\begin{equation}
P_i^{(3\omega)}=\epsilon_0 \Sigma_{jkl}\chi_{ijkl}E_jE_kE_l
\end{equation}
where $\chi_{ijkl}$ is the fourth-rank nonlinear susceptibility tensor. 

The contracted form of the tensor for monoclinic and triclinic crystals, respectively, is given by\cite{yang1995expression}:
\begin{widetext}
\begin{equation}
\begin{bmatrix}
\chi_{11} & 0 & \chi_{13} & \chi_{14} & \chi_{15} & \chi_{16} & \chi_{17} & \chi_{18} & 0 & 0 \\
0 & \chi_{22} & 0 & \chi_{24} & 0 & 0 & 0 & 0 & \chi_{29} & \chi_{20} \\
\chi_{31} & 0 & \chi_{33} & 0 & \chi_{35} & \chi_{36} & \chi_{37} & \chi_{38} & 0 & 0 \\
\end{bmatrix}
\end{equation}

\begin{equation}
\begin{bmatrix}
\chi_{11} & \chi_{12} & \chi_{13} & \chi_{14} & \chi_{15} & \chi_{16} & \chi_{17} & \chi_{18} & \chi_{19} & \chi_{10} \\
\chi_{21} & \chi_{22} & \chi_{23} & \chi_{24} & \chi_{25} & \chi_{26} & \chi_{27} & \chi_{28} & \chi_{29} & \chi_{20} \\
\chi_{31} & \chi_{32} & \chi_{33} & \chi_{34} & \chi_{35} & \chi_{36} & \chi_{37} & \chi_{38} & \chi_{39} & \chi_{30} \\
\end{bmatrix}
\end{equation}

\end{widetext}

For in-plane excitation, only the first two rows of each tensor are relevant and the dominant tensor components are: 
\begin{align}
\chi_{11}&=\chi_{xxxx}^{(3)}, 
\chi_{12}=\chi_{xyyy}^{(3)}, 
\chi_{18}=\chi_{xxyy}^{(3)},
\chi_{19}=\chi_{xxxy}^{(3)},\\ \nonumber
\chi_{21}&=\chi_{yxxx}^{(3)},
\chi_{22}=\chi_{yyyy}^{(3)}, \chi_{28}=\chi_{yxyy}^{(3)}, 
\chi_{29}=\chi_{yxxy}^{(3)}.
\end{align}

For linearly polarized excitation the electric field lies in the $x$-$y$ plane and is expressed as $\mathbf{E}(\theta) = E_0(\cos\theta\,\hat{x} + \sin\theta\,\hat{y})$, where $\theta$ is the polarization angle measured from the $x$ ($b$) axis.

This leads to the following measurable intensities for monoclinic and triclinic crystals, respectively:

\begin{align}
I_x^{(3\omega)}(\theta) &\propto (\chi_{11}\cos^3(\theta)+3\chi_{18}\cos(\theta)\sin^2(\theta))^2\\
I_y^{(3\omega)}(\theta) &\propto (\chi_{22}\sin^3(\theta)+3\chi_{29}\sin(\theta)\cos^2(\theta))^2\\
I_x^{(3\omega)}(\theta) &\propto (\chi_{11}\cos^3(\theta)+\chi_{12}\sin^3(\theta)\label{tri_Ix}\\
&+3\chi_{18}\cos(\theta)\sin^2(\theta)+3\chi_{19}\sin(\theta)\cos^2(\theta))^2, \nonumber\\
I_y^{(3\omega)}(\theta) &\propto (\chi_{21}\cos^3(\theta)+\chi_{22}\sin^3(\theta)\label{tri_Iy}\\
&+3\chi_{28}\cos(\theta)\sin^2(\theta)+3\chi_{29}\sin(\theta)\cos^2(\theta))^2. \nonumber
\end{align}

\begin{figure}
\includegraphics[width=0.5\textwidth]{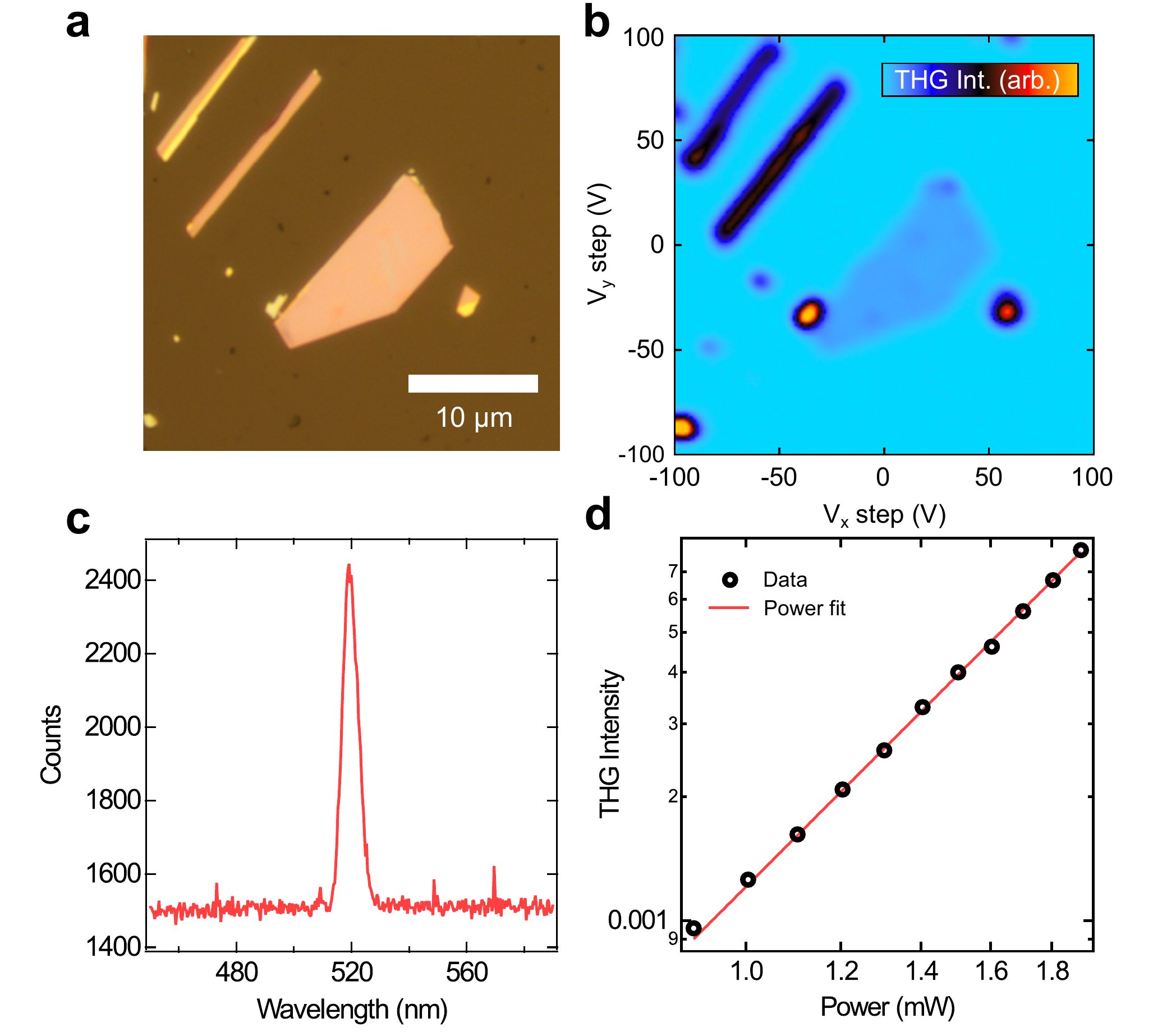}
\caption{\label{fig2} Third harmonic generation in TMT crystals. (a) Optical microscopy image of exfoliated TiS$_3$ flakes. (b) THG intensity scan of the third harmonic for the same region shown in panel (a). (c) Spectrum of the third harmonic from a representative TiS$_3$ flake. (d) THG intensity as a function of laser power for a representative TiS$_3$ flake.}
\end{figure}

We plot $I_x$, $I_y$ for monoclinic and triclinic crystals in Figure 1. Figure 1a plots the third harmonic intensity for a monoclinic crystal with all susceptibility values equal to 1 ($\chi_{11}=\chi_{18}=\chi_{22}=\chi_{29}=1$). For these values, sampling the signal at fixed crystallographic axes results in symmetric butterfly-like patterns for the $x$ and $y$ components. Figure 1b shows the third harmonic intensity as a function of polarization for a triclinic crystal with all susceptibility coefficients equal to 1 ($\chi_{11}=\chi_{12}=\chi_{18}=\chi_{19}=\chi_{21}=\chi_{22}=\chi_{28}=\chi_{29}=1$). For this case, the $x$ and $y$ components are equal for triclinic crystals (Eqns. \ref{tri_Ix}, and \ref{tri_Iy}). A predominantly two-lobe pattern is observed with an angle of 45$^{\circ}$ relative to the TMT $b$ axis (x axis) of the crystal. For crystals with strong anisotropies, the susceptibility coefficients will differ greatly. For example, Figure 1c plots the $x$ and $y$ components for a monoclinic crystal with the following susceptibility coefficients: $\chi_{11}=1, \chi_{18}=0.1$ (black horizontal two-lobe pattern) and $\chi_{22}=0.5, \chi_{29}=0.1$ (red vertical two-lobe pattern). Anisotropy in the susceptibility coefficients also greatly changes the expected triclinic third harmonic. Figure 1d plots the third harmonic intensity for a triclinic crystal with a single susceptibility coefficient of unity for each component and all others suppressed. The black two-lobe pattern is observed for $\chi_{11}=\chi_{22}=1$ and $\chi_{12}=\chi_{18}=\chi_{19}=\chi_{21}=\chi_{28}=\chi_{29}=0.1$ while the red four-lobe pattern is observed for $\chi_{18}=\chi_{28}=1$ and $\chi_{11}=\chi_{12}=\chi_{19}=\chi_{21}=\chi_{22}=\chi_{29}=0.1$. 

\begin{figure*}
\includegraphics[width=\textwidth]{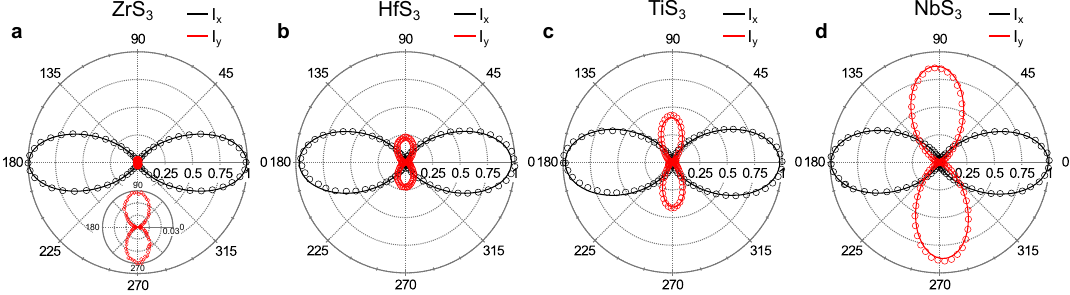}
\caption{Polarization-dependent third harmonic generation for selected flakes of each TMT crystal. Each panel shows the normalized THG intensity as a function of excitation polarization angle for the b-axis analyzer (black open circles, $I_x$) and the a-axis analyzer (red open circles, $I_y$), along with the corresponding tensor fits (solid lines). Data are normalized to the maximum of the b-axis channel. (a) ZrS$_3$ (flake 1A), monoclinic, $\chi_{11}/\chi_{22} = 5.82$. Inset shows a zoom of the a-axis data. (b) HfS$_3$ (flake 3B), monoclinic, $\chi_{11}/\chi_{22} = 2.08$. (c) TiS$_3$ (flake 2F), monoclinic, $\chi_{11}/\chi_{22} = 1.55$. (d) NbS$_3$ (flake 1A), triclinic, $\chi_{11}/\chi_{22} = 1.06$.}
\label{fig3}
\end{figure*}

\section{\label{sec:level1}Isolation and characterization of TMTs}

Bulk TMT material is obtained through solid--gas reaction between pure metals and sulfur powder sealed in a glass ampule and heated to a temperature of 550 $^{\circ}$C for 60 hours\cite{flores2016synthesis}. The powders containing bulk-like nanowires are characterized using x-ray diffraction (XRD) to determine the crystal structure (see Supplemental Material below\nocite{autere,qfls,tpa,qiu,tse,dusabe,pdpse,maragkakis2024}). ZrS$_3$, HfS$_3$, and TiS$_3$ all present XRD patterns that are consistent with monoclinic structures while the NbS$_3$ crystallizes in the triclinic crystal system. In order to isolate single flakes, the powders are  placed on scotch tape or polydimethylsiloxane (PDMS) stamps to facilitate exfoliation onto quartz substrates (0.5 mm thick). Figure 2a shows an optical microscopy image of a representative TiS$_3$ flake on a quartz substrate. Scanning the excitation laser over the surface of the sample and collecting the third harmonic at 520 nm produces a THG intensity map as shown in Figure 2b displaying features that closely match those found in the microscopy image (Figure 2a). The measured spectrum of the reflected light around the wavelengths expected for the third harmonic shows a narrow peak at 520 nm (Figure 2c). This signal has a cubic dependence on the incident laser power (Figure 2d) consistent with third harmonic generation. The fit line (red) extracts a power-law exponent of 2.90 ± 0.03. The cubic power dependence shown in Fig. 2d is a representative measurement confirming the third-order nature of the signal; in the remaining materials the third harmonic was identified by its emission at one-third of the excitation wavelength (520~nm) and its rejection of the fundamental by the bandpass filter.The thickness of the flakes is measured using a Park Instruments atomic force microscope (AFM). 

Third harmonic generation in TMT crystals is measured using a custom microscope (see the Supplemental Material below for details). A femtosecond laser (100 fs pulse width) is directed through a Glan--Taylor polarizer and a half-wave plate to condition and control the polarization of the incident beam. A pair of short-pass dichroic mirrors directs the laser through the objective and toward the sample; the two are oriented so that the polarization-dependent
phase and reflectance of the second mirror are compensated by the first, preserving a clean linear polarization state at the sample. The back-scattered light is collected with the same objective, passes back through the second dichroic and an analyzer, and is directed through a bandpass filter to an avalanche photodiode.

Figure \ref{fig3} shows the polarization-dependent THG response for representative flakes of all four TMT crystals. In each panel, the b-axis analyzer signal ($I_x$, black open circles) and the a-axis analyzer signal ($I_y$, red open circles) are plotted as a function of excitation polarization angle and overlaid with fits to the symmetry-allowed tensor equations (Eqs. 5-8). All datasets are normalized to the maximum of the b-axis channel and rotated so that the dominant $\chi_{11}$ lobe is aligned along the 0$^\circ$/180$^\circ$ axis. For the monoclinic crystals this dominant lobe coincides with the long, needle-like axis of the flake, which is the crystallographic b-axis (chain direction) of the TMTs. We verified this correspondence optically by comparing the flake morphology, imaged in situ with the integrated camera, against the known orientation of the incident linear polarization, so that the axis assignment is fixed by the flake morphology rather than inferred from the third harmonic response alone. For thin NbS$_3$ the in-plane anisotropy is too weak for the THG lobes to define an axis, and the b-axis is taken from the flake morphology. The extracted tensor components are reported normalized to $\chi_{11}$, the dominant component along the b-axis, so that all reported coefficients are dimensionless ratios referenced to that flake's own b-axis response; the anisotropy ratio is independent of this choice and serves as our primary, normalization-independent metric for comparison across flakes and materials. These reported ratios are effective values that include
linear-optical (Fresnel and dichroic) factors; we estimate these factors in the
Supplemental Material below and find that they are modest and act to increase the
intrinsic anisotropy.

The three monoclinic crystals, ZrS$_3$, HfS$_3$, and TiS$_3$, are each well described by Eqs. 5 and 6.  In all three cases the b-axis channel ($I_x$) exhibits a strongly two-lobe pattern dominated by $\chi_{11}$, while the a-axis channel ($I_y$) is described primarily by $\chi_{22}$ and $\chi_{29}$. ZrS$_3$ displays the largest anisotropy ratio, $\chi_{11}/\chi_{22} = 5.82$, consistent with the giant anisotropy previously reported for this material \cite{zhu2025orientation}. The a-axis signal for ZrS$_3$ flake 1A is very small and plotted again in the inset of Figure 3a. HfS$_3$ and TiS$_3$ show progressively reduced anisotropy ratios of 2.08 and 1.55, respectively, with both analyzer channels cleanly resolved and well described by the monoclinic model.

NbS$_3$ is the only triclinic member of the family investigated here and is therefore fit using the more general expressions in
Eqs. 7 and 8, which include the additional
off-diagonal tensor components $\chi_{12}$, $\chi_{19}$, $\chi_{21}$, and $\chi_{28}$. The lower triclinic symmetry of NbS$_3$ permits additional independent tensor components, and accordingly both the off-diagonal terms and the full pair of channels contribute to the fit. For this representative flake the measured anisotropy ratio is small ($\chi_{11}/\chi_{22} = 1.06$), with both lobes of comparable amplitude. 

\begin{figure*}
\includegraphics[width=\textwidth]{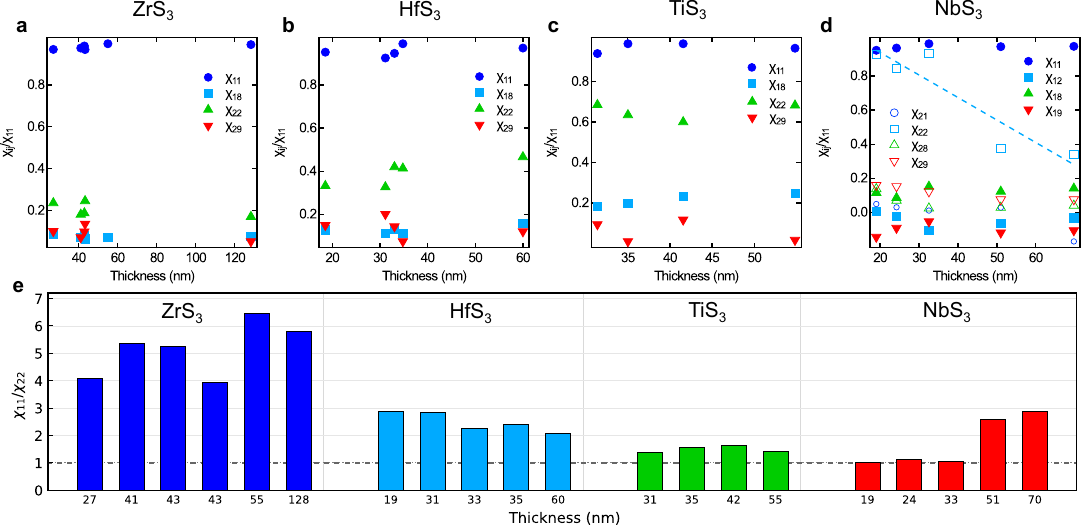}
\caption{Extracted third-order nonlinear susceptibility tensor components and anisotropy ratios for all measured flakes of the four TMT crystals. (a--d) Normalized tensor components $\chi_{ij}/\chi_{11}$ as a function of flake thickness for (a) ZrS$_3$, (b) HfS$_3$, (c) TiS$_3$, and (d) NbS$_3$. Filled symbols denote the dominant in-plane components ($\chi_{11}$, $\chi_{18}$, $\chi_{22}$, $\chi_{29}$) for the monoclinic crystals. In panel (d)  filled symbols denote ($\chi_{11}$, $\chi_{12}$, $\chi_{18}$, $\chi_{19}$) and open symbols denote ($\chi_{21}$, $\chi_{22}$, $\chi_{28}$, $\chi_{29}$). All components are normalized to $\chi_{11}$. The dashed line in panel (d) is a linear guide to the eye for $\chi_{22}$, which decreases with increasing thickness. (e) Anisotropy ratio $\chi_{11}/\chi_{22}$ for every flake, grouped by material and ordered by thickness within each group; the thickness of each flake (in nm) is indicated below each bar. The dotted line marks the equal-amplitude condition $\chi_{11}/\chi_{22} = 1$. The anisotropy ratio decreases systematically across the family, from ZrS$_3$ through HfS$_3$ and TiS$_3$, approaching unity for the thinner NbS$_3$ flakes.}
\label{fig4}
\end{figure*}

The polarization-dependent third harmonic measurements were repeated for multiple exfoliated flakes of each crystal spanning a range of thicknesses, and the in-plane susceptibility tensor components were extracted from fits to Eqs. 5-8. See Table~\ref{tab:components} in the Supplemental Material below for a summary of all measured flakes. Figure \ref{fig4}a-d shows the extracted components as a function of thickness. For the three monoclinic crystals, the response is dominated by the diagonal components $\chi_{11}$ and $\chi_{22}$, with the off-diagonal terms $\chi_{18}$ and $\chi_{29}$ remaining small ($\lesssim 0.2$) across all thicknesses measured. Within the thickness range investigated ($\sim$19--128 nm), the normalized components show no strong systematic thickness dependence for ZrS$_3$, HfS$_3$, and TiS$_3$, indicating that the anisotropic nonlinear response in this regime is governed by the intrinsic crystal symmetry rather than by layer number. NbS$_3$, the only triclinic member, requires the additional off-diagonal components $\chi_{12}$, $\chi_{19}$, $\chi_{21}$, and $\chi_{28}$, whose small but nonzero values produce the lobe asymmetry characteristic of the lower crystal symmetry. In contrast to the monoclinic crystals, NbS$_3$ exhibits a pronounced thickness dependence: $\chi_{22}/\chi_{11}$ increases toward unity as the flake thickness decreases, so that the two polarization channels become comparable in amplitude and the anisotropy is strongly reduced in the thinnest flakes. 

This trend is summarized in Fig. \ref{fig4}(e), which plots the anisotropy ratio $\chi_{11}/\chi_{22}$ for every flake measured. The ratio is largest for ZrS$_3$ (reaching $\chi_{11}/\chi_{22} \approx 6.5$) and decreases systematically across the family through HfS$_3$ ($\chi_{11}/\chi_{22} \approx 2$--$3$) and TiS$_3$ ($\chi_{11}/\chi_{22} \approx 1.4$--$1.6$). Our largest measured amplitude ratio, $\chi_{11}/\chi_{22} \approx 6.5$, is in reasonable agreement with the value reported by Zhu \textit{et al.} once the difference in convention is accounted for, their intensity ratio of 61.5 corresponds to an amplitude ratio of $\sqrt{61.5} \approx 7.8$. Zhu et al. further report that this anisotropy varies with flake thickness, in apparent contrast to the thickness-independent ratios we observe for the monoclinic crystals. We attribute this difference to the excitation conditions of the two experiments. Zhu et al. excite at 1400 to 1500 nm, placing the third harmonic at 2.48 to 2.66 eV, at or above the strongly b-polarized absorption edge of ZrS$_3$, whereas our 1560 nm excitation places the third harmonic at 2.38 eV, nearer the onset of that edge.

We place the third-order response for these materials on an absolute scale by referencing the third harmonic signal of each flake to monolayer WSe$_2$ and MoS$_2$ measured under identical conditions. The extracted $b$-axis susceptibilities are of order $1\times10^{-19}$~m$^2$/V$^2$ for TiS$_3$, ZrS$_3$, and NbS$_3$, comparable to those of monolayer transition metal dichalcogenides at the same wavelength; a full account, including the comparison in Table~S2 and Figure S3, is given in the Supplemental Material below.

\section{\label{sec:level1}Discussion}

The systematic decrease in anisotropy across the monoclinic members is consistent with the proximity of the third harmonic photon energy ($3\hbar\omega \approx 2.4$~eV) to each material's strongly $b$-polarized absorption edge (Fig.~\ref{fig:schematic}). For TiS$_3$ this picture is supported directly by the recent wavelength-dependent measurements of Zhang \textit{et al.}\cite{zhang2026efficient}, who observe a $b$-polarized THG resonance near 1585~nm and compute a $b$-polarized absorption that exceeds the $a$-polarized response across 1.80--2.70~eV. Their HSE06 calculation places the indirect gap of TiS$_3$ at 1.06~eV, so that $3\hbar\omega$ lies well above the edge, in a spectral region where the $a$-polarized absorption has also become appreciable. The activation of this second channel raises $\chi_{22}$ relative to the chain-polarized $\chi_{11}$ and compresses the anisotropy ratio, accounting for the small $\chi_{11}/\chi_{22}$ of TiS$_3$. For ZrS$_3$ and HfS$_3$, whose larger gaps ($\sim$1.9~eV)\cite{island2017electronics} place $3\hbar\omega$ just above the band edge, the $a$-polarized channel remains weak and the chain-polarized response dominates; the wavelength-dependent THG of ZrS$_3$ reported by Zhu \textit{et al.}\cite{zhu2025orientation} shows an analogous resonant enhancement near the band edge. Because the gaps of ZrS$_3$ and HfS$_3$ are nearly equal, the larger ratio of ZrS$_3$ cannot be attributed to resonance proximity alone, and more likely reflects a strongly $b$-polarized band-edge transition governed by the differing $4d$ (Zr) and $5d$ (Hf) metal--sulfur hybridization\cite{msulfur}.

NbS$_3$ stands apart from the monoclinic members. Its anisotropy ratio falls from $\chi_{11}/\chi_{22} \approx 2.6$--$2.9$ in the thickest flakes ($\gtrsim 50$ nm) toward unity below $\sim$25 nm, opposite to the confinement-driven enhancement of the THG anisotropy reported for monoclinic ZrS$_3$, whose anisotropy ratio increases as the flakes are thinned\cite{zhu2025orientation} , pointing to a mechanism specific to this crystal. Unlike the other members, triclinic NbS$_3$ (NbS$_3$-I) owes its reduced symmetry to a Peierls-type Nb--Nb dimerization along the $b$-axis chains \cite{nbs3nanowire}. Removing this dimerization yields a non-dimerized $P2_1/m$ structure essentially identical to ZrS$_3$, favored by only $\sim$0.18 eV per unit cell \cite{nbs3nanowire}. Because such Peierls distortions are strain-tunable \cite{peierlsstrain, tmtstrain} and exfoliated flakes carry inhomogeneous substrate strain that is largest in the thinnest layers, we hypothesize that partial relaxation of the dimerization drives the thinnest flakes toward the undimerized, ZrS$_3$-like ($P2_1/m$) limit, consistent with the anisotropic strain response of the related compound ZrSe$_3$ \cite{zrse3strain}. We note that the substrate strain in the thinnest flakes could also influence the anisotropy through routes not tied to the dimerization, and we cannot fully separate such a contribution from the partial relaxation proposed here.

\section{Conclusion}

We have investigated the polarization-dependent third harmonic generation of four transition metal trichalcogenides, ZrS$_3$, HfS$_3$, TiS$_3$, and NbS$_3$, using non-resonant excitation at 1560~nm. By fitting the polarization dependence along the crystallographic a- and b-axes to the symmetry-allowed forms of the fourth-rank nonlinear susceptibility tensor, we extracted the dominant in-plane components for multiple flakes of each material and quantified the anisotropy through the ratio $\chi_{11}/\chi_{22}$. The ratios are largest for ZrS$_3$ (reaching $\approx 6.5$) and decrease across the family through HfS$_3$ and TiS$_3$. We associate this trend with the proximity of the third harmonic energy ($3\hbar\omega \approx 2.4$~eV) to the strongly $b$-polarized absorption edge of each material. NbS$_3$ presents pronounced anisotropy in thick flakes but its anisotropy is strongly reduced in the thinnest flakes, opposite to the confinement-driven trend of the monoclinic crystals, which we attribute to a partial relaxation of the Peierls-type Nb--Nb dimerization that drives the thinnest flakes toward the undimerized, ZrS$_3$-like limit.

These results establish the transition metal trichalcogenides as a family of 2D materials with large, symmetry-controlled tunable nonlinear optical anisotropy, making them an attractive platform for polarization-sensitive nonlinear photonic applications such as on-chip frequency conversion, polarization-resolved imaging, and anisotropy-based optical components. Future work combining wavelength-dependent measurements and
first-principles calculations of the band-edge transitions
would clarify the microscopic origin of the nonlinear anisotropy across the
family.

\begin{acknowledgments}
This work was supported by the National Science Foundation under Grant No.~2047509. Research was sponsored by the Army Research Office and was accomplished under Grant No.~W911NF-23-1-0160. The views and conclusions contained in this document are those of the authors and should not be interpreted as representing the official policies, either expressed or implied, of the Army Research Office or the U.S. Government. The U.S. Government is authorized to reproduce and distribute reprints for Government purposes notwithstanding any copyright notation herein. E.F., J.R.A., and I.J.F. thank the financial support of the Spanish MICINN under the project PID2021-126098OB-I00.
\end{acknowledgments}

\section*{Data Availability}
The data that support the findings of this article are not publicly available. The data are available from the authors upon reasonable request.


\bibliography{TMT_bib}

\clearpage
\onecolumngrid
\begin{center}
\textbf{\large Supplemental Material: Third harmonic generation in the transition metal trichalcogenides}
\end{center}
\vspace{1em}
\twocolumngrid

\setcounter{section}{0}
\setcounter{equation}{0}
\setcounter{figure}{0}
\setcounter{table}{0}
\renewcommand{\thesection}{S\Roman{section}}
\renewcommand{\theequation}{S\arabic{equation}}
\renewcommand{\thefigure}{S\arabic{figure}}
\renewcommand{\thetable}{S\arabic{table}}

\section{X-ray diffraction for as-grown TMTs}
\begin{figure}
\includegraphics[width=0.5\textwidth]{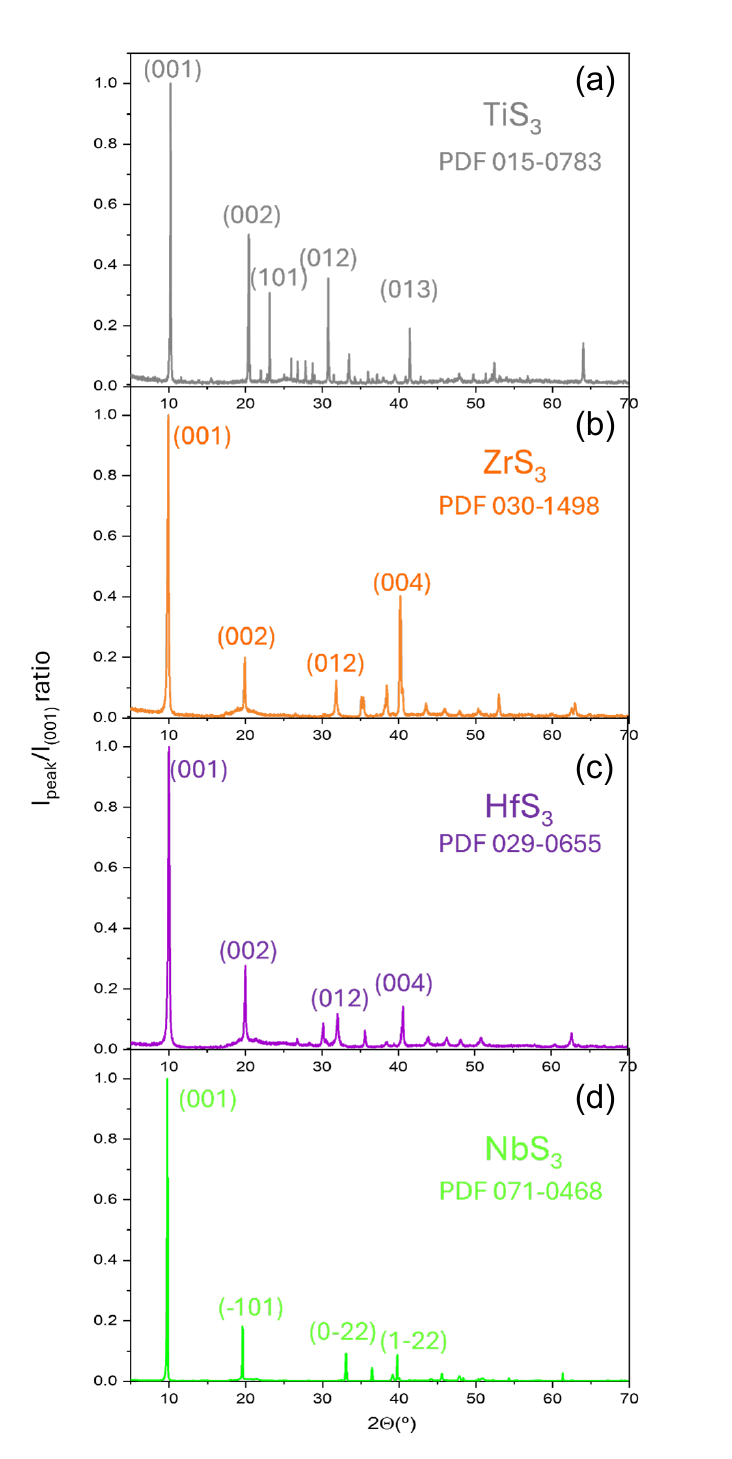}
\caption{\label{figs1} X-ray diffraction patterns for all four materials studied including TiS3 (a), ZrS3 (b), HfS3 (c), NbS3 (d). The patterns are consistent with monoclinic structure for TiS3, ZrS3, and HfS3 and triclinic structure for NbS3.   }
\end{figure}

The crystallographic structures of the as-grown powders are identified using x-ray diffraction (XRD). Figure \ref{figs1} shows the XRD patterns for TiS3 (a), ZrS3 (b), HfS3 (c), and NbS3 (d). The first three materials (a-c) are identified as having monoclinic crystal structures according to JCPDS-ICDD Powder Diffraction Files (PDF) 015-0783 (TiS3), 030-1498 (ZrS3), and 029-0655 (HfS3). NbS3 is identified as triclinic according to PDF 071-0468.

\section{Custom scanning microscope setup}

\begin{figure*}
\includegraphics[width=\textwidth]{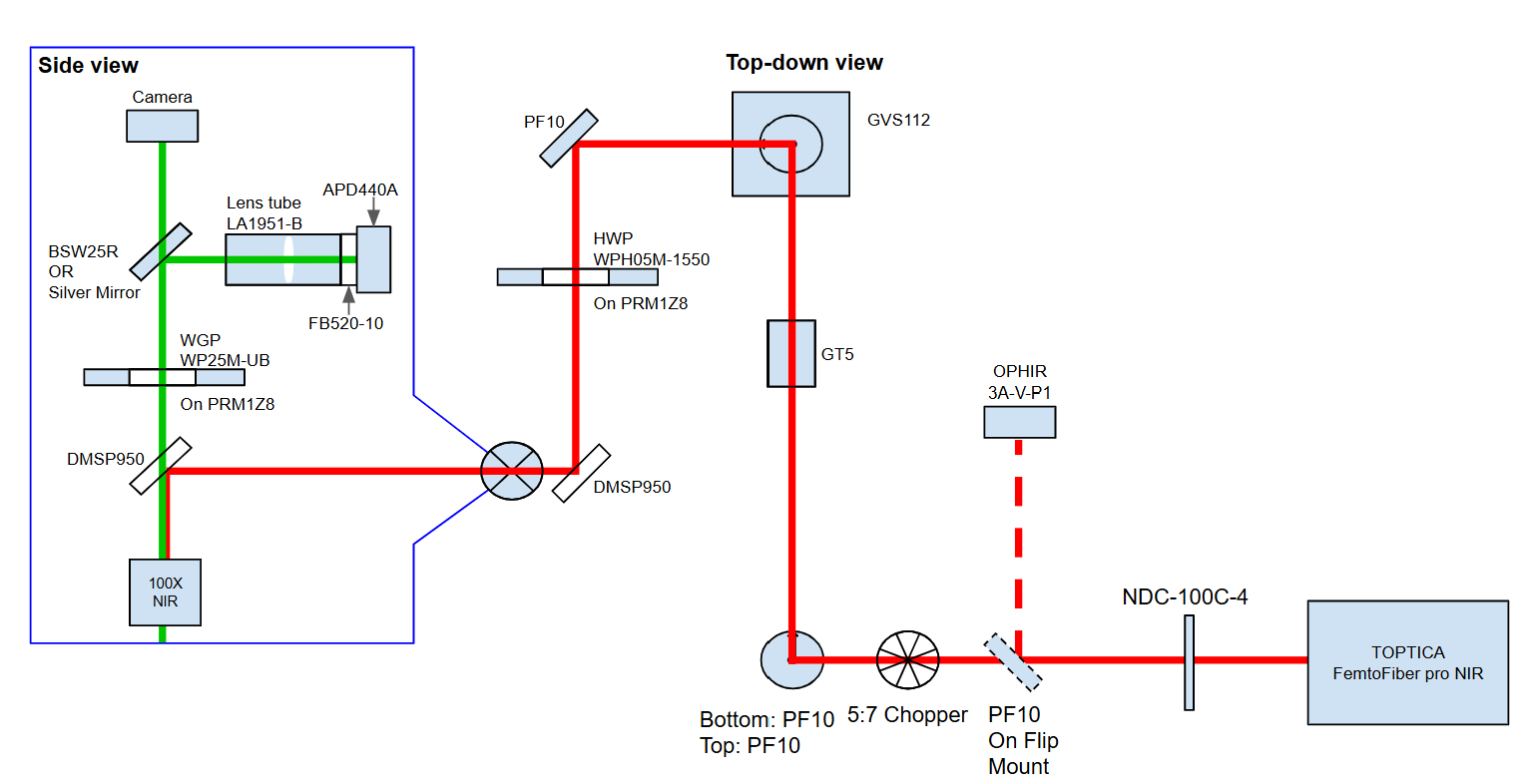}
\caption{\label{figs2}Custom scanning microscopy setup.}
\end{figure*}

A schematic of the custom confocal microscope used for the third harmonic generation measurements is shown in Fig.~\ref{figs2}, with a top-down view of the excitation path and a side view of the microscope column. The excitation source is a TOPTICA FemtoFiber pro NIR fiber laser ($\sim$100~fs pulse width, 1560~nm center wavelength). The output passes first through a continuously variable neutral-density filter (Thorlabs NDC-100C-4) used to set the incident power. A flip-mounted pickoff mirror (PF10) can be inserted into the beam to divert it to a thermal power meter (OPHIR 3A-V-P1) for in-situ measurement of the incident power; with the mirror removed, the beam continues along the excitation path.

The beam is then modulated by a 5:7 optical chopper at 357 Hz to enable lock-in detection of the third harmonic signal, and is steered by a series of protected-silver mirrors (PF10). A Glan-Taylor calcite polarizer (GT5) defines a clean linear polarization state, and a half-wave plate (HWP, Thorlabs WPH05M-1550) mounted on a motorized rotation stage (PRM1Z8) controls the orientation of the incident linear polarization. Rotating the half-wave plate by an angle $\theta_{\mathrm{HWP}}$ rotates the polarization by $2\theta_{\mathrm{HWP}}$; the polarization-dependent data are reported as a function of this doubled angle. A pair of galvanometer scanning mirrors (Thorlabs GVS112) raster the beam across the sample to acquire the topographic third harmonic maps.

As shown in the side view, the conditioned beam is directed into the microscope column by a pair of short-pass dichroic mirrors (Thorlabs DMSP950), which reflect the 1560 nm fundamental while transmitting the visible third harmonic near 520 nm. Two identical dichroic mirrors are used rather than one: reflection from a dichroic mirror at non-normal incidence introduces uncontrolled wavelength dependent differences in reflectance and phase shift between the s- and p-polarized components of the incident beam, which would alter the polarization state of the reflected beam and corrupt the polarization-dependent measurements. The two mirrors are oriented with their planes of incidence perpendicular, which swaps the s and p designation of the components, so that the reflectance and phase differences for the first reflection are canceled by the equal and opposite differences for the second reflection, preserving a clean linear polarization for the beam steered into the microscope objective. The two mirrors were obtained from a single coating batch through Thorlabs to make their reflectance and phase characteristics as nearly identical as possible. The beam is then focused onto the sample by a 100$\times$ near-infrared objective. The back-scattered light is collected by the same objective and passes back through the microscope dichroic, which separates the harmonics from the reflected fundamental.

For polarization-resolved detection, a wire-grid polarizer (WGP, Thorlabs WP25M-UB) mounted on a motorized rotation stage (PRM1Z8) selects the analyzed polarization component of the third harmonic; the a-axis, b-axis, and total (no analyzer) configurations referred to in the main text correspond to the settings of this analyzer. The third harmonic is then directed to the detection optics by a silver mirror, or alternatively a beamsplitter (Thorlabs BSW25R) which may be used to visually inspect the flake and harmonic signal. The signal passes through a lens tube to reduce background light and passes through a lens (Thorlabs LA1951-B) and bandpass filter (Thorlabs FB520-10) centered at 520~nm to reject any residual fundamental and stray light before reaching an avalanche photodiode (Thorlabs APD440A). The photodiode output is recovered with a lock-in amplifier referenced to the chopper frequency. A camera,
accessed by switching the beamsplitter into the path may be used in tandem with a white light source that is directed into the optical path via a flip-mount mirror placed between the dichroic pair for wide-field optical imaging of the sample for locating and aligning individual flakes. 

\section{Sample overview}

Table~\ref{tab:components} summarizes the in-plane third-order susceptibility components and anisotropy ratios extracted for every flake measured in this study. For each flake, the polarization-dependent third harmonic intensity was recorded in the $a$-axis and $b$-axis analyzer channels and fit to the symmetry-allowed tensor expressions of the main text: Eqs.~(5) and~(6) for the monoclinic crystals (ZrS$_3$, HfS$_3$, TiS$_3$) and Eqs.~(7) and~(8) for the triclinic crystal (NbS$_3$). Because the measurement records only the third harmonic intensity, the fitted coefficients are real effective amplitudes and do not carry the phase of the complex susceptibility. Each set of components is normalized to $\chi_{11}$, the dominant component along the $b$-axis, so that all tabulated values are dimensionless ratios referenced to the $b$-axis response of the same flake. The anisotropy ratio $\chi_{11}/\chi_{22}$ is independent of this normalization and is used throughout as the primary metric for comparison across flakes and materials.

For the three monoclinic crystals the response is governed by the diagonal components $\chi_{11}$ and $\chi_{22}$, with the off-diagonal terms $\chi_{18}$ and $\chi_{29}$ remaining small compared with the dominant diagonal components at all thicknesses; the components $\chi_{12}$, $\chi_{19}$, $\chi_{21}$, and $\chi_{28}$ vanish by the monoclinic mirror symmetry and are omitted from the fit. The triclinic crystal NbS$_3$ requires the full set of off-diagonal components, whose small but nonzero values produce the lobe asymmetry characteristic of the lower crystal symmetry. Within each material the flakes are ordered by thickness and all thicknesses were measured using an Atomic Force Microscope (AFM). The anisotropy ratio decreases systematically across the family, from ZrS$_3$ through HfS$_3$ to TiS$_3$, while NbS$_3$ shows a pronounced thickness dependence, with $\chi_{11}/\chi_{22}$ falling toward unity (equal-amplitude lobes) in the thinnest flakes. The representative flakes shown in Fig.~3 of the main text are ZrS$_3$ 1A, HfS$_3$ 3B, TiS$_3$ 2F, and NbS$_3$ 1A.

\begin{table*}
\centering
\caption{\label{tab:components} Normalized in-plane third-order susceptibility
components and anisotropy ratio for every measured flake. All components are
normalized to the $b$-axis response and are therefore dimensionless. The
anisotropy ratio is reported as $\chi_{11}/\chi_{22}$ (b-axis/a-axis). The
monoclinic crystals (ZrS$_3$, HfS$_3$, TiS$_3$) are fit with Eqs.~(5)--(6) in the main text; the
triclinic crystal (NbS$_3$) requires the additional off-diagonal components
$\chi_{12}$, $\chi_{19}$, $\chi_{21}$, $\chi_{28}$ of Eqs.~(7)--(8). Note that $\chi_{22}$ and $\chi_{29}$ values for ZrS3, flake 6B were taken from a total signal fit because the a-axis data is near our noise floor.}
\footnotesize
\setlength{\tabcolsep}{6pt}
\begin{tabular}{llcccccccccc}
\hline\hline
Material & Flake & $t$ (nm) & $\chi_{11}$ & $\chi_{12}$ & $\chi_{18}$ & $\chi_{19}$ & $\chi_{21}$ & $\chi_{22}$ & $\chi_{28}$ & $\chi_{29}$ & $\chi_{11}/\chi_{22}$ \\
\hline
ZrS$_3$ & 1D (up) &  27.0 & 0.970 &        & 0.086 &        &        & 0.236 &        & 0.101 & 4.10 \\
ZrS$_3$ & 1C      &  41.1 & 0.976 &        & 0.073 &        &        & 0.182 &        & 0.071 & 5.37 \\
ZrS$_3$ & 4B      &  42.9 & 0.986 &        & 0.079 &        &        & 0.188 &        & 0.100 & 5.26 \\
ZrS$_3$ & 1D (dn) &  43.3 & 0.969 &        & 0.065 &        &        & 0.246 &        & 0.136 & 3.94 \\
ZrS$_3$ & 6B      &  55.0 & 0.997 &        & 0.072 &        &        & 0.154 &        & 0.007 & 6.46 \\
ZrS$_3$ & 1A      & 128.3 & 0.993 &        & 0.077 &        &        & 0.171 &        & 0.053 & 5.82 \\
\hline
HfS$_3$ & 2A (dn) &  18.6 & 0.952 &        & 0.128 &        &        & 0.332 &        & 0.150 & 2.86 \\
HfS$_3$ & 2A (up) &  31.1 & 0.925 &        & 0.111 &        &        & 0.327 &        & 0.201 & 2.83 \\
HfS$_3$ & 1J      &  33.0 & 0.946 &        & 0.134 &        &        & 0.420 &        & 0.144 & 2.25 \\
HfS$_3$ & 1A      &  34.8 & 0.991 &        & 0.112 &        &        & 0.413 &        & 0.075 & 2.40 \\
HfS$_3$ & 3B      &  60.0 & 0.971 &        & 0.158 &        &        & 0.466 &        & 0.121 & 2.08 \\
\hline
TiS$_3$ & 2H      &  31.4 & 0.939 &        & 0.183 &        &        & 0.685 &        & 0.094 & 1.37 \\
TiS$_3$ & 2F      &  35.0 & 0.987 &        & 0.199 &        &        & 0.635 &        & 0.010 & 1.56 \\
TiS$_3$ & 1L      &  41.6 & 0.987 &        & 0.234 &        &        & 0.601 &        & 0.117 & 1.64 \\
TiS$_3$ & 1M      &  54.8 & 0.965 &        & 0.248 &        &        & 0.683 &        & 0.017 & 1.41 \\
\hline
NbS$_3$ & 2B-L &  19.0 & 0.951 &  0.006 & 0.116 & -0.144 &  0.049 & 0.926 & 0.138 & 0.160 & 1.03 \\
NbS$_3$ & 2B-R &  24.2 & 0.964 & -0.023 & 0.086 & -0.092 &  0.030 & 0.843 & 0.069 & 0.153 & 1.14 \\
NbS$_3$ & 1A   &  32.5 & 0.990 & -0.107 & 0.152 & -0.054 &  0.011 & 0.932 & 0.024 & 0.123 & 1.06 \\
NbS$_3$ & 2A   &  51.0 & 0.972 & -0.063 & 0.122 & -0.119 &  0.028 & 0.375 & 0.028 & 0.075 & 2.59 \\
NbS$_3$ & 3A   &  69.8 & 0.975 & -0.031 & 0.141 & -0.106 & -0.170 & 0.339 & 0.040 & 0.076 & 2.88 \\
\hline\hline
\end{tabular}
\end{table*}

\section{Absolute third-order susceptibility}

We placed the third-order nonlinear response of the trichalcogenides on an absolute scale by referencing the third-harmonic signal of each flake to monolayer WSe$_2$ and MoS$_2$ measured under identical excitation at 1560~nm. The third-harmonic intensity generated by an absorbing slab of thickness $L$ is given by~\cite{zhu2025orientation}
\begin{equation}
I_{3\omega}(L) \propto L^2\,\lvert\chi^{(3)}\rvert^{2}\,
\frac{\sin^{2}\!\left(\tfrac{\Delta k L}{2}\right)
      + \sinh^{2}\!\left(\tfrac{\alpha L}{4}\right)}
     {\left(\tfrac{\Delta k L}{2}\right)^{2}
      + \left(\tfrac{\alpha L}{4}\right)^{2}}\,
e^{-\alpha L/2}\, I_{\omega}^{3},
\label{eq:slab}
\end{equation}
where $\alpha$ is the absorption coefficient at the third harmonic, $I_\omega$ is the incident intensity, and $\Delta k = k(3\omega) - 3k(\omega) = \tfrac{6\pi}{\lambda}\,(n_{3\omega}+n_{\omega})$ is the phase mismatch in the reflection geometry, with $\lambda$ the fundamental wavelength and $n_\omega$ and $n_{3\omega}$ the refractive indices at the fundamental and third harmonic. Writing the thickness-dependent factor of Eq.~(\ref{eq:slab}) as $G(L)$, the ratio of the third-harmonic signal from a flake to that from a reference material measured at the same incident intensity gives the absolute susceptibility of the dominant $b$-axis component,
\begin{equation}
\lvert\chi^{(3)}_{11}\rvert = \chi^{(3)}_{\mathrm{ref}}
\sqrt{\frac{I_{3\omega}}{I_{3\omega,\mathrm{ref}}}\,
      \frac{G_{\mathrm{ref}}}{G(L)}}\,,
\label{eq:ratio}
\end{equation}
in which the common detection efficiency, collection geometry, and incident intensity cancel. For the monolayer references the thickness factor reduces to $G \approx L^2$. We use $\chi^{(3)}=1.2\times10^{-19}$~m$^2$/V$^2$ and a thickness of 0.7~nm for monolayer WSe$_2$, with its optical constants at 1560~nm ($\alpha=7.5\times10^{4}$~cm$^{-1}$, $n_\omega=3.54$, $n_{3\omega}=4.21$) taken from Zhu \textit{et al.}~\cite{zhu2025orientation}; the susceptibility of monolayer MoS$_2$ was obtained from our WSe$_2$-referenced measurement and is consistent with reported values. This reference value is itself corroborated by the independent measurement of Autere \textit{et al.}~\cite{autere}, who obtained $\chi^{(3)}=1.0\times10^{-19}$~m$^2$/V$^2$ for monolayer WSe$_2$ at the same 1560~nm excitation, in agreement with the value used here to within twenty percent. The two reference monolayers are mutually consistent: extracting each material's susceptibility against WSe$_2$ and against MoS$_2$ independently yields values that agree to within approximately ten percent, and the extraction is reproducible across excitation powers, confirming the absolute scale is not tied to a single reference standard. The monolayer thickness of the WSe$_2$ and MoS$_2$ reference flakes was confirmed both by atomic force microscopy and by optical-contrast analysis on the quartz substrate, compared against reported monolayer contrast for these materials~\cite{qfls,tpa}.

The optical constants used for each material are drawn from the following sources, and their origin (measured, calculated, or extrapolated) is stated explicitly, as it sets the confidence of the corresponding absolute value. For ZrS$_3$, the axis-resolved $n$ and $\kappa$ at both the fundamental (1560~nm) and third-harmonic (520~nm) wavelengths are taken from the first-principles calculations of Qiu \textit{et al.}~\cite{qiu}. For TiS$_3$, the axis-resolved $n$ and $\kappa$ at 520~nm are taken from the experimental micro-reflectance data of Papadopoulos \textit{et al.}~\cite{papadopoulos2018large}, and the refractive index at 1560~nm, which lies below the gap, is taken from the first-principles static value of Tse~\cite{tse}. For NbS$_3$, no experimental optical constants are available; we therefore use the first-principles (DFT and $GW$--BSE) values of Dusabe \textit{et al.}~\cite{dusabe}, computed for the triclinic NbS$_3$-I phase studied here, at both wavelengths. The absolute value for TiS$_3$ therefore rests on measured optical constants at the third harmonic (with a calculated fundamental index), whereas the ZrS$_3$ and NbS$_3$ values rest on calculated constants throughout and carry a correspondingly larger uncertainty. We note, however, that for the thin flakes used here the extracted $\chi^{(3)}$ is only weakly sensitive to the optical constants: recomputing the ZrS$_3$ value with a different set of optical constants changes it by less than one percent. For HfS$_3$, axis-resolved optical constants are not available in the literature, and an absolute susceptibility is therefore not extracted; its anisotropy ratio, which is independent of any optical-constant input, is reported throughout.

\begin{figure*}
\includegraphics[width=\textwidth]{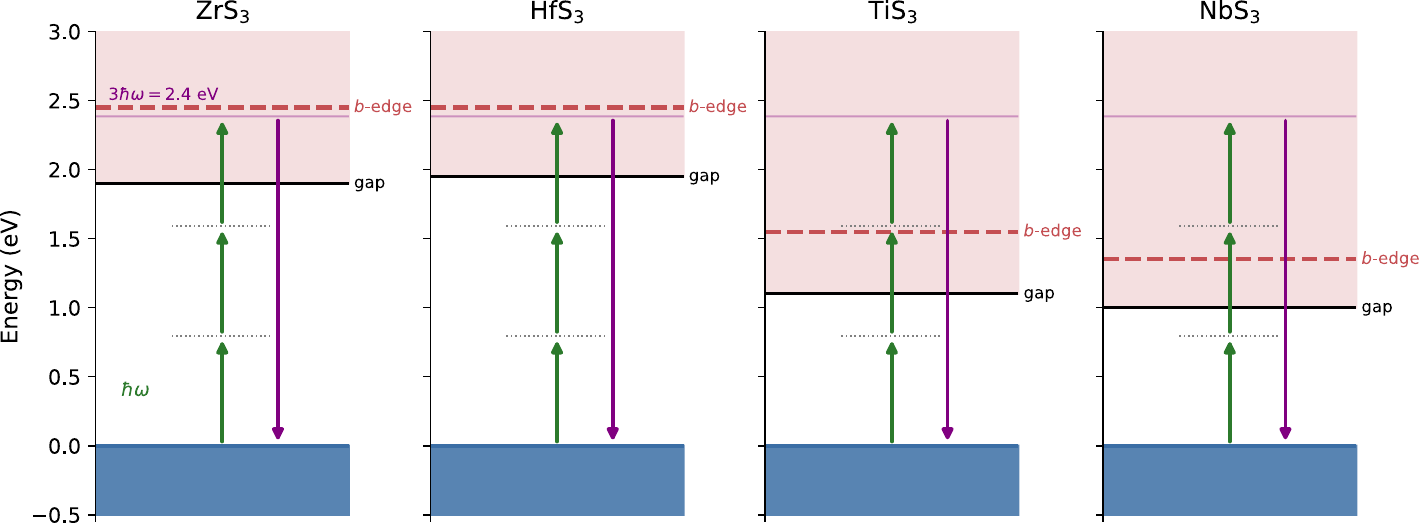}
\caption{\label{fig:schematic} Schematic of the third-harmonic generation
process in the transition metal trichalcogenides, showing the third-harmonic
energy relative to the fundamental gap and the $b$-polarized absorption edge of
each material. For each material the filled valence band (blue) and the
fundamental gap (solid line) are shown together with the $b$-polarized
absorption edge (red dashed line). Three fundamental photons
($\hbar\omega = 0.795$~eV at 1560~nm, green) drive the process through virtual
intermediate levels (dotted), generating a third harmonic at
$3\hbar\omega = 2.4$~eV (purple). For ZrS$_3$ and HfS$_3$ the third harmonic
lies nearer the $b$-polarized absorption edge, where the chain-polarized
($b$-axis) response is more strongly enhanced and the anisotropy is largest,
whereas for TiS$_3$ and NbS$_3$ it lies well above the edge, within the
absorption continuum, where transitions of mixed polarization reduce the net
anisotropy.
The diagram is schematic and the energies are approximate.}
\end{figure*}

The extracted $b$-axis susceptibilities are $\chi_{11}\approx1.5\times10^{-19}$, $1.1\times10^{-19}$, and $1.2\times10^{-19}$~m$^2$/V$^2$ for TiS$_3$, ZrS$_3$, and NbS$_3$, respectively, and are summarized in Table~\ref{tab:chi3comparison} alongside literature values. Three features of this comparison are notable. First, these values lie within the range reported for monolayer transition metal dichalcogenides at the same 1560~nm excitation, $1.0$ to $3.6\times10^{-19}$~m$^2$/V$^2$ for WSe$_2$, MoSe$_2$, WS$_2$, and MoS$_2$~\cite{autere}, and are comparable to those of other anisotropic 2D materials such as PdPSe~\cite{pdpse}. Our susceptibilities are therefore typical of the 2D nonlinear-optical materials and require no special enhancement to explain. Second, the absolute susceptibilities are similar across the trichalcogenide family, whereas the anisotropy ratio $\chi_{11}/\chi_{22}$ varies strongly from material to material; the distinguishing nonlinear-optical property of the trichalcogenides is therefore the large, symmetry-controlled anisotropy of the response rather than the magnitude of the susceptibility itself. Third, our values are smaller than the giant susceptibility reported for ZrS$_3$ by Zhu \textit{et al.}~\cite{zhu2025orientation}, which was obtained under near-resonant excitation at 1400 to 1500~nm, where the third harmonic lies higher on the $b$-polarized absorption edge than at our excitation (Fig.~\ref{fig:schematic}) and the chain-polarized response is correspondingly more enhanced; we note that absolute $\chi^{(3)}$ determinations for 2D materials are known to vary substantially with measurement conditions, including the substrate, which alone can shift the extracted value by several fold~\cite{autere}. Regardless of the magnitude of the susceptibility, the anisotropy we quantify is a robust property of the crystal symmetry.

\begin{table*}
\centering
\caption{\label{tab:chi3comparison} Comparison of the third-order nonlinear
optical response of the transition metal trichalcogenides with reference
monolayers, other monolayer transition metal dichalcogenides, and additional
anisotropic 2D materials. For our materials, $\chi^{(3)}$ is the absolute
magnitude of the dominant $b$-axis ($\chi_{11}$) component, extracted by
referencing the third-harmonic signal to monolayer WSe$_2$ and MoS$_2$ measured
under identical non-resonant excitation at 1560~nm (see text); the anisotropy is
the amplitude ratio $\chi_{11}/\chi_{22}$. Literature values are reported as
given in the cited works; intensity anisotropy ratios (marked $I$) equal the
square of the amplitude ratios (marked $A$) used here.}
\footnotesize
\setlength{\tabcolsep}{5pt}
\begin{tabular}{lllll}
\hline\hline
Material & $\chi^{(3)}$ (m$^2$/V$^2$) & Anisotropy & Gap (eV) & Source / notes \\
\hline
TiS$_3$ & $\sim1.5\times10^{-19}$ & $\chi_{11}/\chi_{22}=1.4$--1.6 & 1.1  & this work; 1560~nm, non-resonant \\
ZrS$_3$ & $\sim1.1\times10^{-19}$ & $\chi_{11}/\chi_{22}=3.9$--6.5 & $\sim$1.9 & this work; 1560~nm, non-resonant \\
NbS$_3$ & $\sim1.2\times10^{-19}$ & $\chi_{11}/\chi_{22}=1.0$--2.9 & $\sim$1.0 & this work; 1560~nm, non-resonant \\
HfS$_3$ & --                      & $\chi_{11}/\chi_{22}=2.1$--2.9 & $\sim$1.9 & this work; absolute value not extracted \\
WSe$_2$ & $1.2\times10^{-19}$ & isotropic & 1.6 & this work (reference); value from Ref.~\cite{zhu2025orientation} \\
MoS$_2$ & $\sim2.5\times10^{-19}$ & isotropic & 1.8 & this work (reference) \\
\hline
ZrS$_3$ & $6.8\times10^{-17}$ & $61.5^{I}$ (17.7~nm) & $\sim$1.9 & Ref.~\cite{zhu2025orientation}; near-resonant, 1400--1500~nm \\
TiS$_3$ & $1.0\times10^{-19}$ & $2.15^{I}$          & 1.06 & Ref.~\cite{zhang2026efficient}; intrinsic, strain-tunable to $317^{I}$ \\
MoS$_2$ & $3.6\times10^{-19}$ & isotropic & 1.8 & Ref.~\cite{autere}; monolayer, 1560~nm \\
MoSe$_2$& $2.2\times10^{-19}$ & isotropic & 1.5 & Ref.~\cite{autere}; monolayer, 1560~nm \\
WS$_2$  & $2.4\times10^{-19}$ & isotropic & 2.0 & Ref.~\cite{autere}; monolayer, 1560~nm \\
WSe$_2$ & $1.0\times10^{-19}$ & isotropic & 1.6 & Ref.~\cite{autere}; monolayer, 1560~nm \\
PdPSe   & $6.2\times10^{-19}$ & anisotropic & $\sim$1.3 & Ref.~\cite{pdpse}; anisotropic 2D \\
GeSe    & not reported & $\chi_{11}/\chi_{22}=1.7$--2.1$^{A}$ & 1.1--1.2 & Ref.~\cite{dasgupta}; 1560~nm \\
SnS     & not reported & $0.75\pm0.22^{I}$ (AC/ZZ) & $\sim$1.1 & Ref.~\cite{maragkakis2024}; 1542~nm \\
\hline\hline
\end{tabular}
\end{table*}

\section{Effect of linear optical anisotropy on the extracted ratios}

The polarization-dependent third-harmonic intensity measured outside the flake
is not determined by the nonlinear susceptibility alone. The fundamental field
must couple into the flake, the generated harmonic must couple out, and both
depend on the linear optical constants along the excitation axis. Because the
trichalcogenides are birefringent and linearly dichroic, these factors differ
between the $a$ and $b$ axes and do not cancel in the ratio of the two
polarization channels, so the reported $\chi_{11}/\chi_{22}$ are effective
ratios that include a linear-optical contribution.

We estimate that contribution explicitly as follows. At normal incidence the
fundamental field amplitude transmitted from air into the flake along axis $i$
is governed by the amplitude transmission coefficient
\begin{equation}
\tau^{\mathrm{in}}_i = \frac{2}{1+n^{\omega}_i},
\end{equation}
where $n^{\omega}_i$ is the refractive index at the fundamental (1560~nm). Since
third-harmonic generation involves three fundamental photons, the induced
nonlinear polarization scales as the cube of the internal field, and therefore
as $(\tau^{\mathrm{in}}_i)^3$. The harmonic field generated inside the flake
then couples out through the flake--air interface with the amplitude
transmission coefficient
\begin{equation}
\tau^{\mathrm{out}}_i = \frac{2\,n^{3\omega}_i}{1+n^{3\omega}_i},
\end{equation}
where $n^{3\omega}_i$ is the refractive index at the third harmonic (520~nm),
and is attenuated by reabsorption over its escape path. Collecting these,
the detected third-harmonic intensity along axis $i$ is
\begin{equation}
I_i \;\propto\; \lvert\chi_{ii}\rvert^{2}\,
\bigl(\tau^{\mathrm{in}}_i\bigr)^{6}\,
\bigl(\tau^{\mathrm{out}}_i\bigr)^{2}\,
e^{-\alpha^{3\omega}_i L_{\mathrm{eff}}},
\label{eq:Ii}
\end{equation}
where $\alpha^{3\omega}_i = 4\pi\kappa^{3\omega}_i/\lambda_{3\omega}$ is the
absorption coefficient at the third harmonic, $\kappa^{3\omega}_i$ the
corresponding extinction coefficient, and $L_{\mathrm{eff}}$ an effective escape
path taken as half the flake thickness. The measured ratio of the two channels
is then
\begin{equation}
\frac{I_b}{I_a}
= \underbrace{\left(\frac{\tau^{\mathrm{in}}_b}{\tau^{\mathrm{in}}_a}\right)^{6}
             \left(\frac{\tau^{\mathrm{out}}_b}{\tau^{\mathrm{out}}_a}\right)^{2}
             e^{-(\alpha^{3\omega}_b-\alpha^{3\omega}_a)L_{\mathrm{eff}}}}
             _{\displaystyle \mathcal{F}_b/\mathcal{F}_a}
\;\left\lvert\frac{\chi_{11}}{\chi_{22}}\right\rvert^{2},
\label{eq:ratiolin}
\end{equation}
so that the intrinsic amplitude ratio is recovered from the measured one as
$(\chi_{11}/\chi_{22})_{\mathrm{intrinsic}}
= (\chi_{11}/\chi_{22})_{\mathrm{meas}}\,
\sqrt{\mathcal{F}_a/\mathcal{F}_b}$. For an isotropic medium
$\mathcal{F}_b=\mathcal{F}_a$ and the correction vanishes.

Evaluating Eq.~(\ref{eq:ratiolin}) with the axis-resolved refractive indices and
extinction coefficients adopted above, the linear-optical prefactor
$\mathcal{F}_b/\mathcal{F}_a$ is of order unity for all three materials,
corresponding to a correction of the amplitude ratio $\chi_{11}/\chi_{22}$ of
roughly 15 to 35\%. In every case the $b$-axis is the more strongly absorbing at
the third harmonic, so reabsorption suppresses the dominant channel and the
linear factors act in the same direction, making the intrinsic anisotropy
somewhat larger than the measured value: the intrinsic amplitude ratio exceeds
the measured one by factors of about 1.2 for ZrS$_3$, 1.1 for TiS$_3$, and 1.4
for NbS$_3$. The reported ratios are therefore conservative, in the sense that
correcting for the linear-optical factors would increase rather than decrease
the anisotropy. In all cases the correction is modest and does not alter the
ordering of the anisotropy across the family, for which ZrS$_3$ remains far more
anisotropic than TiS$_3$ and NbS$_3$.

These are order-of-magnitude estimates. They rest on a single-interface
treatment of the Fresnel factors and on axis-resolved optical constants that are
calculated for ZrS$_3$~\cite{qiu} and NbS$_3$~\cite{dusabe} and, for TiS$_3$,
measured at the third harmonic~\cite{papadopoulos2018large} with a calculated fundamental
index~\cite{tse}. A full account would require a birefringent-slab treatment
with measured optical constants at both wavelengths along both axes. We
therefore report the measured ratios throughout, understood as effective values
in the sense of Eq.~(\ref{eq:ratiolin}), and note that the linear-optical
correction is modest, acts to increase the anisotropy, and preserves all of the
qualitative conclusions of this work.

\end{document}